\documentclass[aps,pra,preprint,superscriptaddress,longbibliography]{revtex4-1}

\usepackage{amsmath}
\usepackage{amsthm}
\usepackage{amsfonts}
\usepackage{amssymb}
\usepackage{xcolor,graphicx}
\usepackage{tikz}
\usepackage{color}
\usepackage[pdftex]{hyperref} 
\usepackage{array}
\usepackage{dsfont}

\begin{document}
	
	\title{Effects of a thermal inversion experiment on STEM students learning and application of damped harmonic motion}

	\author{O. I. Gonz\'alez-Pe\~na}
	\email[e-mail: ]{oig@tec.mx; ogonzalez.pena@gmail.com}
	\affiliation{Tecnologico de Monterrey, Escuela de Ingenier\'ia y Ciencias, Avenida Eugenio Garza Sada 2501, Monterrey, N.L., Mexico, 64849}	
	
	\author{G. Mor\'an-Soto}
	\email[e-mail: ]{gmorans@clemson.edu}
	\affiliation{Department of Basic Sciences, Instituto Tecnol\'ogico de Durango, Blvd. Felipe Pescador 1830, Nueva Vizcaya, Durango, Dgo, Mexico, 34080.}	

	\author{R. Rodr\'iguez-Masegosa}
	\email[e-mail: ]{rrodolfo@tec.mx}
	\affiliation{Tecnologico de Monterrey, Escuela de Ingenier\'ia y Ciencias, Avenida Eugenio Garza Sada 2501, Monterrey, N.L., Mexico, 64849}	

	\author{B. M. Rodr\'iguez-Lara}
	\email[e-mail: ]{bmlara@tec.mx}
	\affiliation{Tecnologico de Monterrey, Escuela de Ingenier\'ia y Ciencias, Avenida Eugenio Garza Sada 2501, Monterrey, N.L., Mexico, 64849}

	\date{\today}
	
\begin{abstract}
	There are diverse teaching methodologies to promote both collaborative and individual work in undergraduate physics courses.
	However, few educational studies seek to understand how students learn and apply new knowledge through open-ended activities that require mathematical modeling and experimentation focused on environmental problems. 
	In this work, we propose a novel home experiment to simulate the dynamics of a particulate under temperature inversion and model it as damped harmonic motion. 
	Twenty six first year students enrolled in STEM majors answered six qualitative questions after designing and developing the experiment. These questions helped analyze the students epistemological beliefs about their learning process of physics topics and its applications. Results showed that this type of open-ended experiments could facilitate the students understanding of physics phenomena.
	In addition, this experiment showed that it could help physics professors to promote students epistemological development by giving their students the opportunity to search for different sources of knowledge and becoming self-learners instead of looking at the professor as the epistemological authority. 
	At the end, students described this activity as a positive experience that helped them realize alternative ways to apply physics topics in different contexts of their environment. 
\end{abstract}
	
	
	\maketitle

\section{Introduction}

Finding strategies to promote student engagement in introductory physics courses is a challenge of our times. 
For instance, something as simple as identifying students preferences for learning physics by demonstrative problem solving on a blackboard or supervised independent collaborative work may improve the learning process \cite{Lavonen2007}.

In a traditional learning environment, introductory physics curricula is usually designed with the laboratory at the core of the learning process. 
It becomes a place to work on learning theoretical concepts as well as carrying on experimentation \cite{Bailey2000,Parra2020}.
This methodology increases student engagement in physics courses and improves conceptual understanding through manipulation of instruments and materials \cite{Thumper2002, Parra2020} to generate and process experimental data \cite{Thomsen1997}.
A well designed experiment could help STEM students develop self-regulated learning strategies \cite{Thumper2002}, giving them the opportunity to build their own conclusions and boost their knowledge about physical phenomena and its interpretation \cite{Roberts1983}.
Self-regulation and motivation is usually driven by epistemic beliefs \cite{Greene2010,Muis2009} that describe the way students think about the nature of knowledge and knowing \cite{Schommer1990}. 

On the other hand, as engagement and cooperation play an essential role of the learning process in physics \cite{Leung2017}, it is important to develop Modeling Instruction (MI) plans to engage students beyond the four walls of an instruction laboratory.
This is why developing plans to improve the academic success of students by boosting their interest and involvement implementing active learning \cite{Dou2018} using passive content \cite{Lenaerts2002} or interactive content \cite{Escalada1998} as well as social media platforms \cite{Gavrin2017} to promote students interaction with their whole learning environment has become quintessential, and a dire necessity as education evolves and distance learning methodologies are more important for students' development. 

Developing better teaching strategies for physics courses, in general, could facilitate STEM students understanding of physics phenomena. 
A better understanding of physics topics could motivate STEM students and professionals to design and develop open-ended challenges aiming to provide solutions to urgent global issues. 
Here, we focus on the issue of air quality control and pollution to build upon the concept of simple and damped harmonic motion applied to the dynamics of particulate dynamics in the atmosphere. 
This issue is specially relevant for us as five Mexican cities (Mexico City, Monterrey, Guadalajara, Toluca and Leon) rank among the 13 cities with worst air quality according to a recent report from the Organization for Economic Co-operation and Development (OECD) \cite{Ales2019}, and it is common to observe pollutants trapped by temperature inversion in our daily life.
Studying simple and damped harmonic motion applied to the dynamics of a particulate in the atmosphere could provide a fertile ground to boost students cognitive process and to create awareness within the frame of the 2030 Agenda of Sustainable Development Goals. In Sec. II, we state this purpose in detail and follow it with a literature review on approaches to teaching the damped harmonic oscillator in Sec. III. We present the details of the theoretical model and experimental setup in Sec. IV and those of our qualitative research is analyzed on the students epistemological beliefs about their learning process in Sec. V and Sec. VI. Finally, we close with our conclusion in Sec. VIII.

\section{Purpose} \label{sec:Purpose}

Students learning new topics usually undergo an epistemological process where they reason about specific information obtained from different sources, then they claim knowledge of these new topics \cite{Barzilai2014}.
Kitchener's work suggest that open-ended problems or activities are more likely to engage students on an epistemological process than solving problems in class \cite{Kitchener1983}.
Our research aims to help physics educators by analyzing the effect of constructing and experimenting with a simulator of thermal inversion on students' understanding of the damped harmonic oscillator.

In this research, we present a methodology to construct an isobaric troposphere simulator using air confined by a glass vessel where it is possible to introduce a foreign gas and make it oscillate by controlling the temperatures at the bottom and top ends of the container to simulate temperature inversion. 
We present this activity to STEM students enrolled in first year introductory physics courses and ask them to collect experimental data of the dynamics to compare it with a numerical simulation of the damped harmonic oscillator.
Students' goal is to find values for the various parameters of the system that provide a good fit between experiment and theory.
This is a collaborative activity for teams of four that submit a single project report. 
Individuals undergo an argumentative test that serves as evidence to evaluate and accredit the understanding and mastering of a technical competence.
In addition to this academic evaluation process, a cohort of students answered a questionnaire looking for their point of view on the effect of this experiment in their understanding of the thermal inversion phenomenon and its relation to damped harmonic oscillation. 

\section{Literature review} \label{sec:LitRev}

The harmonic oscillator is at the core of our modeling of real-world devices involving integrated circuits, fluid mechanics, optical systems, and quantum technologies among others.
Thus, engaging STEM students in an cognitive process that allows them to claim knowledge of this concept and extend its use beyond particular examples becomes a fundamental objective of physics education.

Pendulum and spring-mass systems are the standard textbook example of harmonic motion in introductory physics lectures.
A spring-mass experiment is simple enough to introduce the idea of damped oscillations by measuring the position of the mass \cite{Castro2013,Marinho2016,Hauko2018,Suchatpong2018,Buachoom2019}. 
Conducting experiments with pendulum may improve student satisfaction under self-evaluation of their learning experience and knowledge of the subject \cite{Festiana2019}.
Of course, real world devices are not completely harmonic; both spring-mass systems \cite{Giliberti2014} and pendulum \cite{Pedersen2019} beyond the small displacement or oscillation angle limit, in that order, serve as examples of basic non-linear models that undergraduate students may build using simple materials.

In addition to physical concepts, the oscillation of complex systems, like membranes or strings, allows the introduction of differential equations \cite{Filipponi2010} and Fourier methods to physics problems \cite{Belozerova1999}.
In this direction, coupling a pair of simple harmonic oscillators may ease introducing the idea of coupled differential equations to students with some experience in classical mechanics \cite{Dolfo2018}.
From the simple to the complex, the harmonic oscillator offers an opportunity to understand the significance of mathematical modeling and visualize the effect of variable manipulation to engage STEM student into learning by physical interpretation \cite{Franklin2013}.

Furthermore, analogies to the physical concept of harmonic oscillators may help understanding the workings of real-world devices and phenomena.
For example, the infrared spectrophotometer \cite{Wright2019} may be modeled as a spring-mass oscillator and the membrane vibration happening inside a microphone \cite{Kraftmakher2009} is an analogy to a driven harmonic oscillator \cite{Fay2000,Kharkongor2018}.
For more advanced courses, the idea of a classical harmonic oscillator may be extended to the quantum realm, for example, using basic calculus and algebra \cite{Borghi2017} or studying fluorescence in diatomic sulfide \cite{Boye2019}. 
The motion of an electron in the presence of a two-dimensional potential is another simple example of harmonic motion \cite{Yancey2009} and analogies using spring-systems with coupled masses may help introducing the formation of quantum bands to STEM students \cite{Roberts2018}.

On the education side, the idea of interactive conceptual instruction using collaborative problem solving \cite{Malik2019}, computational simulations \cite{Maulidina2019} and virtual laboratories \cite{Galan2017}, or alternative learning methods \cite{Tisdell2019} show improvement clarifying misconceptions related to harmonic motion.
Furthermore, flipped learning using software simulation of electrical circuits shows improvement in the students knowledge of the damped harmonic oscillator \cite{Albano2002}.

Our experimental proposal focus on the importance of exploring systems beyond the spring-mass and pendulum in order to boost the epistemic cognitive process. In this direction, electronics present an opportunity to introduce highly controllable damping and non-linearities to harmonic oscillators beyond mechanical systems \cite{Varju2009,Xu2019} and, for advance courses, the ability to produce, for example, time-dependent control to introduce continuous symmetries and its invariants following Noether theorem \cite{Abe2009}.
We may look into the space and introduce the damped harmonic oscillator using the dynamics of particular celestial bodies \cite{Aceves2006} or to more complex setups, for example, lasers and optical resonators \cite{Henningsen2011}, classical gases confined by harmonic potentials \cite{Guery1999}, or the oscillation of a superconductor ring levitated by a magnetic field \cite{Giliberti2018}. In particular, we are interested in fluids as they are exceptionally helpful to elucidate the effect of non-constant friction forces \cite{Shamim2010,Hauko2019}.

In the following, we describe a simple experimental setup that simulates the conditions of temperature inversion and allows STEM students to explore the oscillation of a gas in a controllable analogy of the troposphere.Together with the experiment, a theoretical model was generated. The bibliographic search revealed that modeling the effect of temperature inversion by means of the damped harmonic motion has not been reported; the following experiment also incorporate

\section{Experimental methods} \label{sec:Experiment}

\begin{figure}[htp!]
	\centering	
	\includegraphics{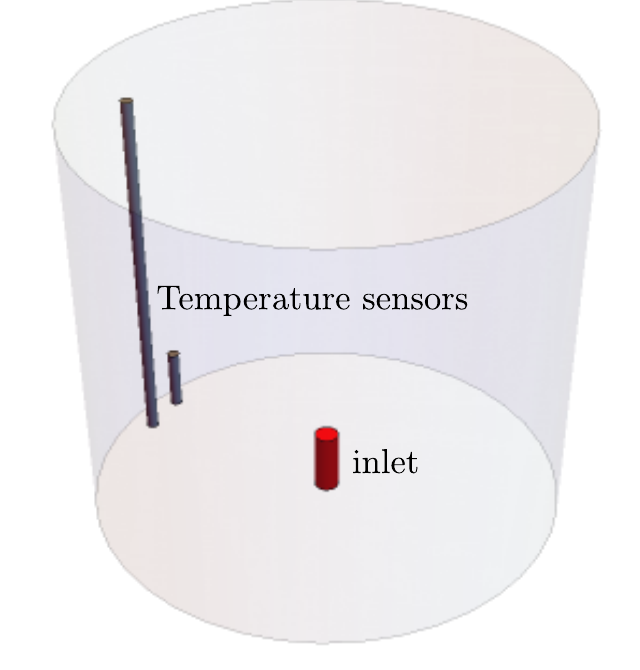}
	\caption{Sketch of the experimental setup used to simulate an isobaric troposphere. A clear container with two temperature sensors and an inlet to inject smoke.} \label{fig:FigSetup}
\end{figure}

The core of our proposal is a toy model of the troposphere to visualize how a gas cloud under temperature inversion behaves like a damped harmonic oscillator.
Our experimental set-up, Fig. \ref{fig:FigSetup}, builds upon the idea that a transparent container whose ends are covered by a good thermal conductor may serve as a simulation of an isobaric troposphere at atmospheric pressure $p$ with control of bottom and top temperatures, $T_{b}$ and $T_{t}$. 
An inlet in the bottom allows us to introduce an external gas that we model as a collection of non-interacting microscopic spheres with constant density $\rho_{g}$ in order to follow its dynamics.

Let us start with the model. 
Temperature control at the bottom and top ends of our simulation allows us to assume an air density that depends on the height. 
For the sake of simplicity, we suppose constant pressure and linear temperature gradient that allows us to approximate the air density,
\begin{align}
\rho_{\mathrm{air}}(y) = \rho_{b} \left( 1 - c\, y\right), \qquad \mathrm{with} \qquad c = \frac{1}{h} \left( 1 - \frac{T_{b}}{T_{t}} \right),
\end{align}
up to first order on the height $y$, that we take as zero at the container bottom and $h$ at the top such that $y \in [0,h]$.
We define the air density at the bottom and top ends, 
\begin{align}
\rho_{b} \equiv \rho_{\mathrm{air}}(0)= \frac{p M}{ R T_{b}} \qquad \mathrm{and} \qquad \rho_{t} \equiv \rho_{\mathrm{air}}(h) = \frac{p M}{ R T_{t}},  
\end{align}
in that order, where our local atmospheric pressure at $540 ~\mathrm{m}$ above the sea level is $p = 101388 ~\mathrm{Pa}$ \cite{presion2020}, the molar mass of air is $M = 28965.4 \times 10^{-6} ~\mathrm{kg~mol}^{-1}$ \cite{molecularmass2004,molecularmass2015}, the ideal gas constant is $R = 8.31447 ~\mathrm{J~mol}^{-1}~\mathrm{K}^{-1}$, and the temperatures are given in Kelvin.

Assuming the foreign gas as a collection of non-interacting microscopic spherical particles, allows us to model the dynamics of each particle using Newton second law,
\begin{align}
m \frac{d^{2} y}{dt^{2}} = -w + F_{B} - b \frac{d y}{d t},
\end{align}
where the forces in the right-hand-side of the equation are the weight $w = m g$ pointing downwards, the buoyant force $F_{B} = \rho_{\mathrm{air}} V_{\mathrm{gas}} g$ pointing upwards, and the Stokes drag for a sphere moving through a viscous fluid proportional to the drag coefficient, $b = 6 \pi r_{\mathrm{gas}} \eta_{\mathrm{air}}$ in terms of the radius of the spherical particle $r_{\mathrm{gas}}$ and the viscosity of the air $\eta_{\mathrm{air}}$, and the velocity of the particle $dy/dt$.

The mass of each gas particle is given by its density and volume, $m = \rho_{\mathrm{gas}} V_{\mathrm{gas}}$, such that its weight becomes $w = \rho_{\mathrm{gas}} V_{\mathrm{gas}} g$ and the dynamics reduce to the following second order differential equation,
\begin{align}
\frac{d^{2} y}{ d t^{2}} = \left[ \frac{\rho_{\mathrm{air}}(y)}{\rho_{\mathrm{gas}}} - 1 \right] g - \frac{b}{\rho_{\mathrm{gas}} V_{\mathrm{gas}}} \frac{d y}{d t},
\end{align}
where we assumed that the gas density change induced by the temperature gradient is negligible at the time scale of the experiment and that the height of the container is small enough to produce no significative changes in the value of the gravity.
For the sake of simplicity, we assume that the viscosity of air has a negligible change with the temperature gradient.
We use our linear approximation to the air density inside the container to unfold the model,
\begin{align}
\frac{d^{2} y}{ d t^{2}} = -\frac{c g}{\rho_{\mathrm{gas}}} \, y + \left( \frac{\rho_{b}}{\rho_{\mathrm{gas}}} - 1 \right) g - \frac{ 9 \eta_{\mathrm{air}}}{ 2 \rho_{\mathrm{gas}} r_{\mathrm{gas}}^{2}}  \frac{d y}{d t},
\end{align}
into that of a damped oscillator, 
\begin{align}
\frac{d^{2} y}{ d t^{2}} = - \omega_{0}^2 y + a_{0} - \gamma \frac{d y}{d t},
\end{align}
where the temperature difference controls the sign of the frequency,
\begin{align}
\omega_{0}^{2} =  \frac{g}{h\rho_{\mathrm{gas}}} \left( 1 - \frac{T_{b}}{T_{t}} \right).
\end{align}
Without considering the rest of the terms in the right-hand-side of the oscillator equation, if the temperature at the top is lower than that at the bottom, $T_{t} < T_{b}$, we have a positive squared frequency $\omega_{0}^{2} > 0$ that yields a harmonic oscillator and we will see our gas sample oscillate. 
In the opposite case, $T_{t} > T_{b}$, we have a negative squared frequency $\omega_{0}^{2} > 0$ that yields an inverted oscillator and our gas sample will rise and remain at the top of the container.
For an isothermal simulation, $T_{t} = T_{b}$ the squared frequency is null, $\omega_{0}^{2} = 0$ and only the external effective acceleration,
\begin{align}
a_{0} = \left( \frac{\rho_{b}}{\rho_{\mathrm{gas}}} - 1 \right) g
\end{align}
has an effect on the dynamics. 
Without considering the rest of the terms in the right-hand-side of the equation, if the gas density is larger than that of the air at the bottom of our simulator, $\rho_{\mathrm{gas}} > \rho_{b}$, the effective external acceleration is negative, $a_{0}<0$, and the gas sinks to the bottom of the container and stay there. 
If it is smaller, $\rho_{\mathrm{gas}} < \rho_{b}$, the effective external acceleration is positive, $a_{0}<0$, and the gas rises to the top of the container and stay there.
If they are equal, $\rho_{\mathrm{gas}} = \rho_{b}$, the effective external acceleration is null and the gas does not move upwards nor downwards.
Finally, the approximate drag frequency for the spherical particles of gas,
\begin{align}
\gamma = \frac{ 9 \eta_{\mathrm{air}}}{ 2 \rho_{\mathrm{gas}} r_{\mathrm{gas}}^{2}},
\end{align}
where we assume a constant air viscosity as it changes from $\eta_{\mathrm{air}}(T=213.5 ~\mathrm{K}) = 0.0171 \times 10^{-3} ~\mathrm{Pa~s}$  to $\eta_{\mathrm{air}}(T=313.5 ~\mathrm{K}) = 0.0218 \times 10^{-3} ~\mathrm{Pa~s}$ for a temperature gradient of $100~\mathrm{{K}}$ \cite{viscosity2003}. 
We take the average of these values as our constant viscosity, $\eta_{\mathrm{air}} = 0.01945\times 10^{-3} ~\mathrm{Pa~s}$ \cite{viscosity2003}.  

Our toy isobaric troposphere model allows the simulation of diverse dynamical phenomena.
Temperature control at the ends of the simulator, for example, allows to switch the driven and damped oscillator between inverted or harmonic behaviour.
Changing the atmospheric or external gases gives even more options to control and explore the parameters of the model.
In the following, we focus our observations on temperature inversion. 

Under standard conditions, the temperature gradually falls with the increase of altitude,
\begin{align}
\Gamma = - \frac{d T}{ d y}.
\end{align}
This is known as the thermal lapse rate; for example, the dry adiabatic lapse rate is around $\Gamma \approx 9.8 \times 10^{-3} ~\mathrm{K ~m}^{-1}$.
Temperature inversion is the phenomenon that occurs when the thermal lapse rate $\Gamma$ changes sign from positive to negative; that is, a hot layer of air with low density hovers above a colder one with high density.
In these situations, it is possible to observe smoke, or other pollutant gases, form a ceiling as the top low density layer of air stops their ascend.

\begin{figure}[htp!]
	\centering	
	\includegraphics{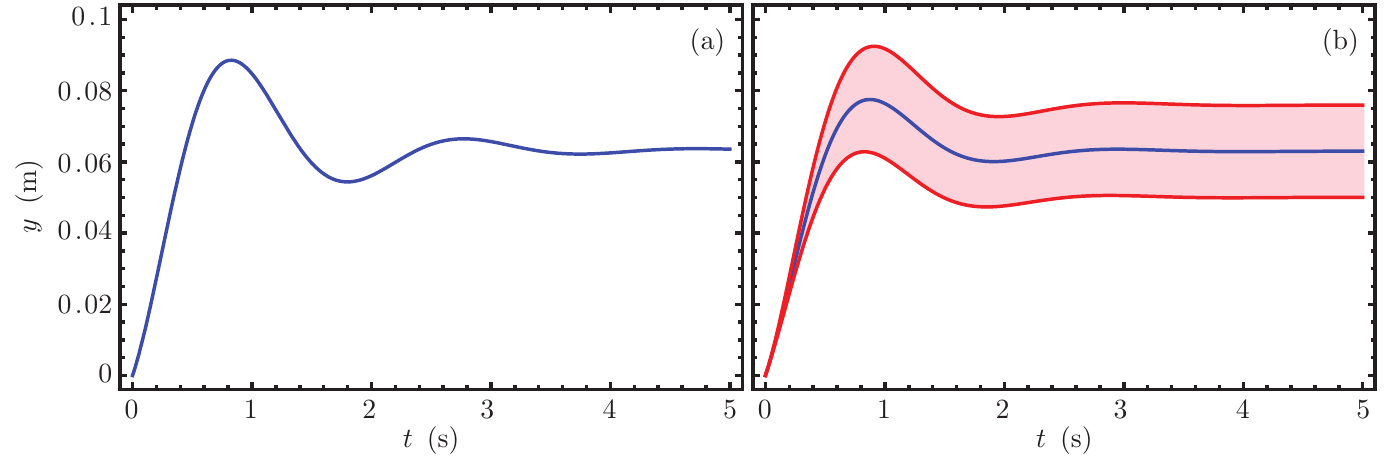}
	\caption{Damped oscillation for (a) a single particle of gas in the experiment with effective parameters provided by Eq. (\ref{eq:Eqw0}) to Eq. (\ref{eq:Eqgamma}) and (b) mean value and region delimited by one standard deviation above and below the mean for a thousand random realizations with parameters following independent normal distributions.} \label{fig:TheoMeas}
\end{figure}

In order to have a reference for the behavior, we present a numerical experiment in a container that is $0.15~\mathrm{m}$ tall, take the standard value of gravity $g = 9.81~\mathrm{m~s}^{-2}$, the bottom of the container at room temperature $T_{b} = 298~\mathrm{K}$ leading to the density of air $\rho_{b} = 1.225~\mathrm{kg~m}^{-3}$ \cite{density2003}, the top of the container heated to $T_{t} = 373.15~\mathrm{K}$ and a constant viscosity of air $\eta_{\mathrm{air}} = 1.945 \times 10^{-5} ~\mathrm{Pa~s}$. 
We assign the gas a density of $\rho_{g} = 1.140~\mathrm{kg~m}^{-3}$ with a particulate radius of $r_{g} = 6 \times 10^{-3}~\mathrm{m}$.
These assumptions provide constants for the differential equation, 
\begin{align}
	\omega_{0} =&~ 3.400 ~\mathrm{rad ~s}^{-1}, \label{eq:Eqw0}\\
	a_{0} =&~ 0.731 ~\mathrm{m ~s}^{-2}, \\
	\gamma =&~ 1.567 ~\mathrm{rad ~s}^{-1} \label{eq:Eqgamma},
\end{align}
leading to the damped behaviour shown in Fig. \ref{fig:TheoMeas}(a). 
Figure \ref{fig:TheoMeas}(b) shows the effect of random variations on the temperatures, densities, effective particulate radius, air viscosity and initial velocity, $\left\{ T_{b}, T_{t}, \rho_{b}, \rho_{\mathrm{gas}}, r_{\mathrm{gas}}, \eta_{\mathrm{air}}, v_{0} \right\}$, following a normal distribution with mean value provided by the parameters above and standard deviation equal to one percent of the mean.
We want to stress how such a small change in parameters produces a strong change in the dynamics.

We ask our students to reproduce the experimental setup, Fig. \ref{fig:FigSetup}, at home using a transparent tempered glass container to avoid fractures from temperature gradients; for instance, we use a coffee jar.
In order to record the temperature at the bottom and top, we use two Vernier Stainless Steel Temperature Probes placed inside the glass container and a Vernier LabQuest Mini controller. 
These may be substituted by simple atmospheric thermometers in contact with the external facet of the container at home.
The sensors and a pewter straw to let smoke in are secured in place on the container lid using modeling clay to guarantee a good seal.
We recommend securing a disposable plate to the container in order to hold ice cubes and secure access to the inlet straw that connects to a smoke container using plastic tubes; for instance, we used a candy jar to contain smoke from paper combustion but a party balloon or plastic bag may play the role. 
A light bulb lamp or an iron covered in aluminium foil may be used to change the temperature at the top of the container.
Finally, we follow smoke dynamics using a logitech HD Pro Webcam C920 and Vernier Logger Pro but any given video capturing device and open source software like Tracker Video Analysis and Modeling Tool should do.

\begin{figure}[htp!]
	\centering	
	\includegraphics{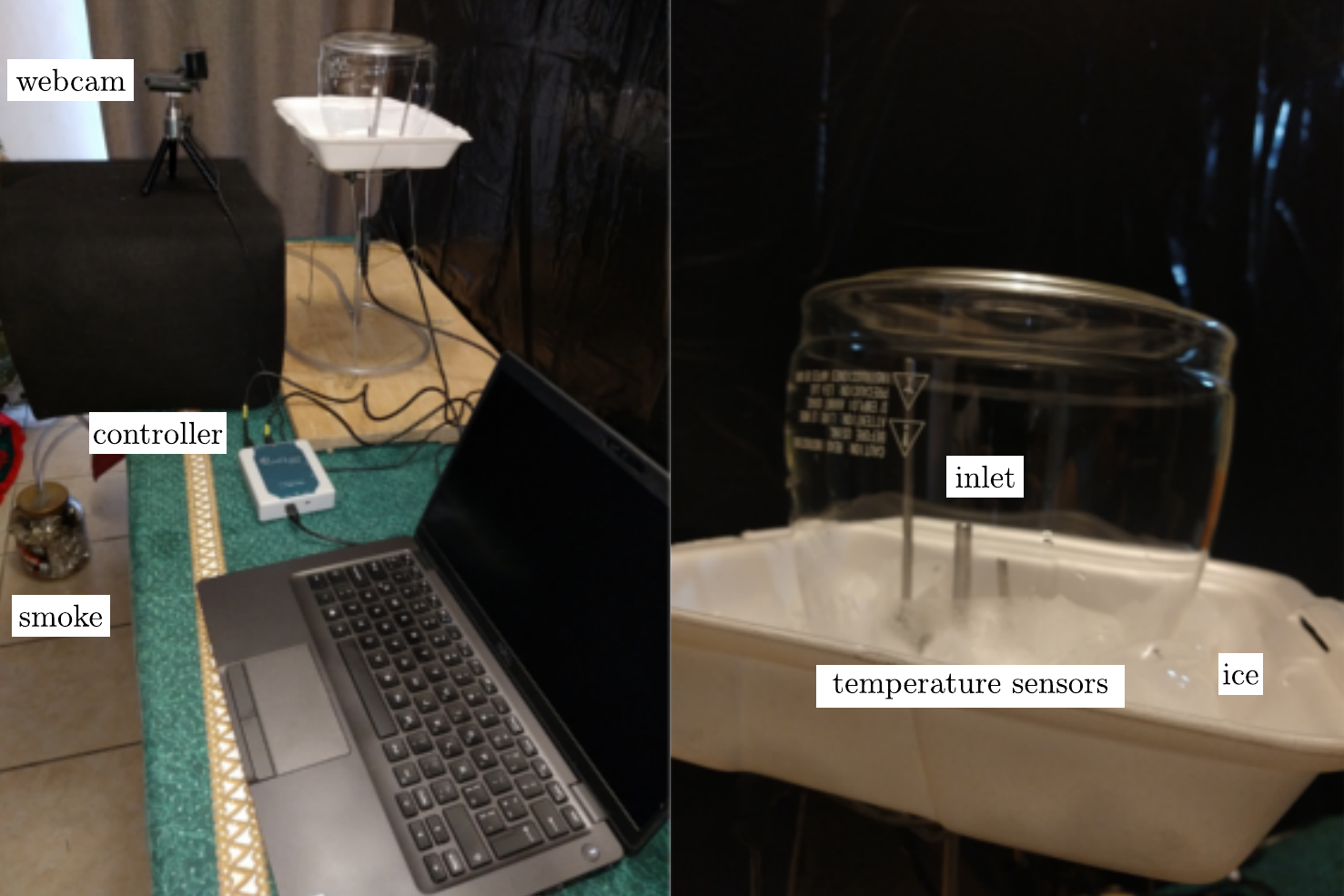}
	\caption{Experimental setup used to simulate temperature inversion in an isobaric troposphere.} \label{fig:FigExperiment}
\end{figure}

\begin{figure}[htp!]
	\centering	
	\includegraphics{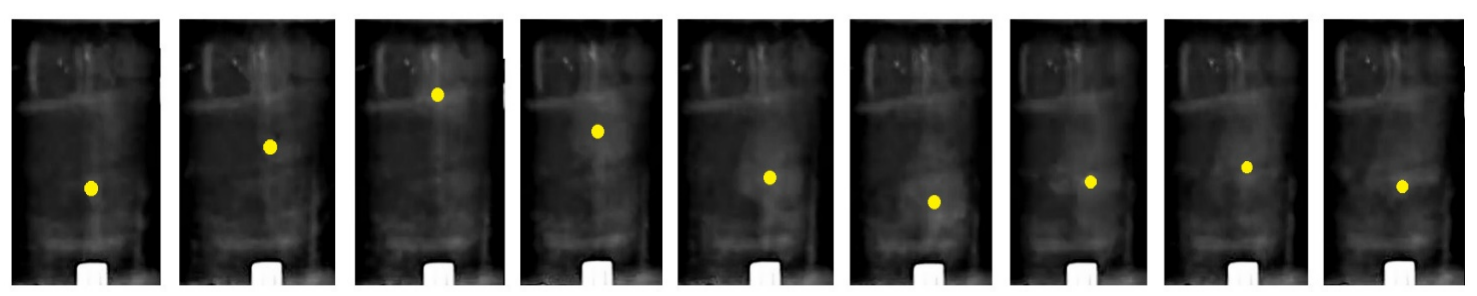}
	\caption{Example from an experimental measurement sequence. The yellow dot indicates the approximated center of mass of the smoke cloud.} \label{fig:MeasSeq}
\end{figure}

Figure \ref{fig:FigExperiment} shows our experimental setup at home.
As the container is not hermetically sealed, the pressure inside should be constant and equal to the atmospheric one.
Our experiment allows controlling the input speed of the smoke as well as the bottom and top temperatures. 
We ask the students to experiment with these three parameters to explore the different dynamical regimes available. 
In particular, we ask for a detailed analysis of a case whose dynamics are an analogy to the damped harmonic oscillator. 
Their experimental data should allow them to fit for the dampening frequency $\gamma$ and the smoke density after figuring out the input velocity of the smoke without any temperature gradient.
Figure \ref{fig:MeasSeq} shows a sequence tracking of the approximated smoke cloud center of mass in an experiment with bottom and top temperatures in the order of $292.35~\mathrm{K}$ and $306.35~\mathrm{K}$, in that order.  

\begin{figure}[htp!]
	\centering	
	\includegraphics{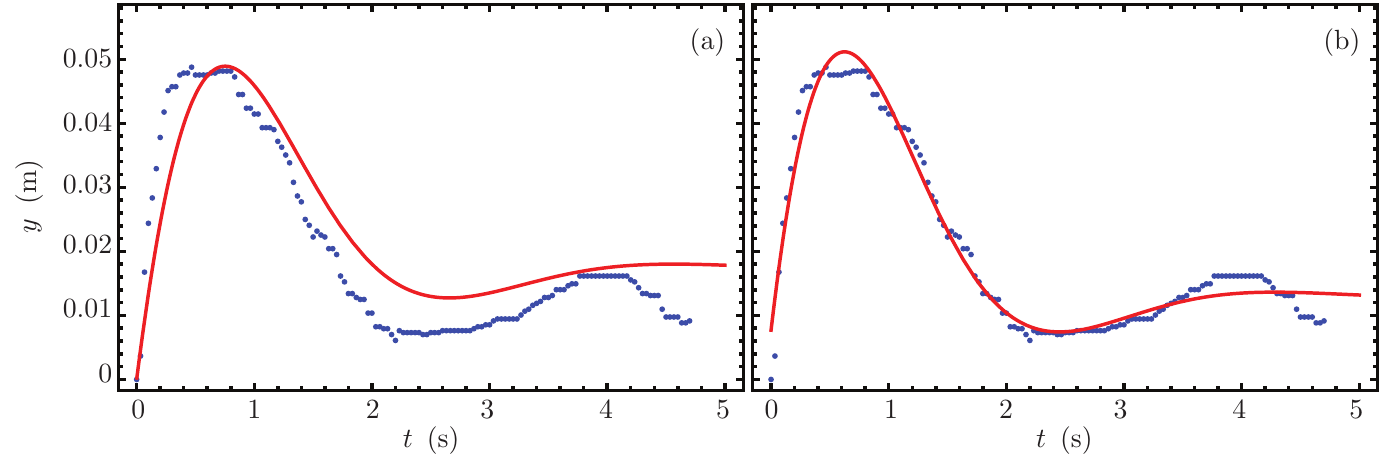}
	\caption{Example from an experimental measurement sequence (data points) and its corresponding fit (solid line) from (a) theoretical parameters and (b) a numerical fit using these parameters as starting point.} \label{fig:ExFit}
\end{figure}

Figure \ref{fig:ExFit}(a) shows experimental data points compared to the analytic model using an atmospheric pressure of $p= 102300~\mathrm{Pa}$ and temperatures of $T_{b} = 292.75~\mathrm{K}$ and $T_{t} = 303.95~\mathrm{K}$ at the top and bottom of the container, in that order, leading to approximate air densities of $\rho_{b}= 1.218 ~\mathrm{kg~m}^{-3}$ and $\rho_{t}= 1.173 ~\mathrm{kg~m}^{-3}$ accounting to a height difference of $h=0.08~\mathrm{m}$ between the temperature sensors. 
We assume the smoke density about $\rho_{\mathrm{gas}} = 1.210~\mathrm{kg~m}^{-3}$ with radius $r_{\mathrm{gas}} = 5.97 \times 10^{-3}~\mathrm{m}$ and initial velocity of the order of $v_{0} = 0.145~\mathrm{m~s}^{-1}$ leading to an effective acceleration, $a_{0}=0.066~\mathrm{m~s}^{-2}$, as well as effective oscillator and damping frequencies $\omega_{0} = 1.93~\mathrm{rad ~s}^{-1}$ and $\gamma = 2.027~\mathrm{rad ~s}^{-1}$, in that order.
The difference between experimental data and the dynamics under our educated guess may arise from variations in any of the assumptions, as we discussed before, and provides an opportunity for discussion.
Our ansatz provides a good starting point for a better fitting using, for example, Newton least squares method or Levenberg-Marquardt method for nonlinear least squares yield an effective acceleration, $a_{0}=0.053~\mathrm{m~s}^{-2}$, effective oscillator and damping frequencies $\omega_{0} = 2.025~\mathrm{rad ~s}^{-1}$ and $\gamma = 2.137~\mathrm{rad ~s}^{-1}$, in that order, and initial velocity $v_{0}= 0.156 \mathrm{m~s}^{-1}$ that provides a better fit shown in Fig. \ref{fig:ExFit}(b).

\section{Qualitative research methods} \label{sec:QualResMet}

In the following, we address the qualitative methods used to analyze our students point of view regarding the effect of this learning activity on their understanding of the damped harmonic oscillation. 

\subsection{Participants}
   
We selected a cohort of 26 students were selected from a total of 185 students that realized the experiment in different physics modules during the fall 2020 semester. These students were taking the introductory physics module with a professor that voluntarily accepted to distribute a questionnaire with six open-ended questions at the end of the experiment. Students were offered extra credit if they decided to answer the questionnaire, and the 100\% of these 26 students answered the six questions. All students were in first semester of different STEM majors in the Tecnologico de Monterrey. Students in this Mexican university are required to participate in multiple challenges related to real-world issues as part of their professional competencies development \cite{TEC212018, Case2019, Graham2018, Parra2020}. These 26 students were taking a 5 weeks-long project for their introductory physics course at the moment of the data collection. 

\subsection{Data collection}

Although physics students are used to conduct experiments to prove their knowledge, this thermal inversion experiment aims to explore the damped movement phenomena in a unique way that may help students to understand this topic. To be able to analyze the effect of this open-ended experiment on students understanding of the damped harmonic oscillation, we distributed a questionnaire with the students after finalizing the experiment. The six questions from the questionnaire were adapted from the Engineering Related Beliefs Questionnaire (ERBQ) \cite{Yu2011}. We selected the ERBQ a starting point to adapt the questions due to its aim for measuring students' beliefs about the nature of STEM-related knowledge and knowing, and also because this instrument has successfully helped to analyze STEM students’ epistemic process after solving open-ended problems in prior studies \cite{Faber2017}. These six questions were selected due to their relationship with open-ended problem solving aspects of the cognitive process of STEM students. The six questions were translated to Spanish language and then adapted to the physics thermal inversion experiment context to ask students about the certainty and sources of physics knowledge. The questionnaire was back translated to English language by an English professor for this paper (see Appendix \ref{app:Questionnaire}).

\subsection{Data Analysis}

The four researchers involved in this study conducted the qualitative analysis. We analyzed the data collected with the questionnaire using open coding to let emerging codes to stay as close as possible to students’ own words and ideas \cite{Charmaz2006}. The final codes for each student were compared side by side with codes from other students aiming to find similarities that could be coded together into meaning units \cite{Charmaz2008}. This coding philosophy helped us to draw conclusions of how the thermal inversion experiment influenced students’ knowledge and understanding of the damped movement, and also helped us to ensure that these meaning units appropriately reflected students’ responses and feelings about the experiment. Students' comments were translated from Spanish to English language by the same professor that translated the questionnaire to be included in the following section.  

\section{Qualitative results} \label{sec:QualRes}

Half of students (13) reported using a single methodology or theory for designing and developing their thermal inversion experiment, while the other half (13) used a combination of two or more different methodologies or theories for their design. Regardless of whether they used only one methodology or different methodologies, students reported that the most likely starting point for their thermal inversion experiment design was the methodology previously explained by their physics professor. Student A4 noted : ``I thought of some less orthodox methods but I mostly focused on the one taught in the lessons.''

Most students (20) reported that they asked for their professor’s help during the designing and development of their thermal inversion experiment. While almost every student asked for their professor’s help, seventeen students reported that they look for more than one source of information asking other professors, classmates, or family members to confirm and complement the information they know about the damped harmonic oscillation. Student A19 noted: ``We asked the challenge’s instructor for help. However, we needed further help and requested it from friends in a senior class.'' Only one student reported completing the inversion experiment without asking for help.

When students were asked if they searched for additional sources of information to complement what they know about the damped harmonic oscillation, almost all of them (25) decided to look for additional information in different sources outside of what they learn in their course. Most of these students (17) looked for more than one source of external information. The most recurrent sources of external information were videos (15) and websites (13), and another less common sources were textbooks (5) and scientific papers (3). Eight students reported looking for only one external source of information, and only one students decided not to look for external sources of information.   

Almost all the students (24) reported that the thermal inversion experiment helped them to better understand the damped harmonic oscillation. These students stated that this experiment helped them to clarify some doubts about damped harmonic oscillations, while they learned different ways to apply this topic in real life problems. Student A23 noted: ``It was really helpful. Before, I thought that harmonic movement only applied to springs, but now I understand that it also applies to more complex systems such as fluids.'' On the other hand, only two students mentioned that this experiment was confusing and it did not help them to understand the damped harmonic oscillation.
Most of the students (23) stated that the thermal inversion experiment could have different final results that could be valid depending of the methodology used during the design, or some differences in the obtained measurements; and seven of these students argued that their responses need to be supported by theory to be considered as a valid response.

Half of students (13) stated that the thermal inversion experiment might have better results in students’ understanding of the damped harmonic oscillation if the professor could spend extra time explaining the theory, and giving them more details about how to apply this topic to solve different problems in different contexts. This issue was more evident for seven students that mentioned having struggles to answer the argumentative test due to difficulties adapting their knowledge and experiences from the thermal experiment to the test context. Student A11 noted: ``I felt prepared to answer questions related to the challenge, but the argumentative exam had nothing to do with thermal inversion.''

\section{Discussion} \label{sec:Discussion}

Students in this research showed that they normally expect instructions from their professor to design the experiment following what they have previously learned and practiced in their physic course. These students were open to follow different methodologies to design and develop their thermal inversion experiment, but they stated that having guidance from any expert in the damped harmonic oscillation phenomena is the key to be able to success in this type of open-ended problems or experiments. This could be related with a low level of episteological maturity for college level students \cite{Perry1970}, where they need to learn new knowledge and ways to apply this knowledge from an epistemological authority that is considered the only source of reliable information \cite{Schommer1990}. Most of students asked for help during the experiment design and development, showing that they are likely to search for an epistemological authority that could help them to develop their knowledge during the experiment process as well. Students that asked for help reported going directly with their physics professor with specific doubts and questions about their thermal inversion experiment. This behavior pinpoints the importance of physics professors and the information they provide to their students before and during the development of the experiment. 

Students reported to be likely to search additional information sources to analyze what they know and solve some doubts, with the videos and websites being the most common places where students look for more information. Very few students consulted science publications and textbooks to expand their knowledge and solve doubts. This lack of interest on scientific publications could be considered an area of opportunity to improve students’ research abilities, and may be considered by physics professors when they advise their students about the advantages of looking for information supported by scientific evidence.

At the end of the thermal inversion experiment and the argumentative test almost all students described this activity as a positive experience that helped them to better understand the damped harmonic oscillation and how this physic phenomena could be applied in different contexts. This type of open-ended experiments or activities could help students develop their knowledge and make them think in different sources where they can search for information to solve their doubts. Giving students the opportunity to experience these type of experiments could help physics professors facilitate students’ epistemological development. Reaching higher levels of epistemological maturity could benefit students development as critical thinkers, making them more likely to think that their knowledge development is their responsibility and they need to search and confirm their own knowledge and believes \cite{Perry1970}. This is relevant for physics professors because some students stated that they felt that the knowledge about the thermal inversion that they learned and practiced was not transferable to different applications of the damped harmonic oscillation. This lack of abilities to apply the same physics principles to solve different problems is common in the low levels of epistemological maturity \cite{Perry1970}, and this position about how to learn and apply new topics could hinder students’ possibilities to understand complex physics phenomena. 

\section{Conclusion} \label{sec:Conclusion}

It is important that physics professors provide enough information to their students to give them confidence to successfully complete open-ended activities like our thermal inversion experiment. 
The process of preparing students with the basic knowledge to perform experiments like the one presented in this research needs to be conducted with certain precautions. Physics professors should provide enough information to their students without leading the entire experiment and giving students the opportunity of solving possible issues by their own. Physics professors should try to conduct these open-ended activities as facilitators, making their students responsible of solving their own doubts by searching sources of reliable information that might help them to develop their knowledge and solve their struggles. 
Our thermal inversion experiment could be used by physic professors to reinforce the students knowledge about damped harmonic oscillations, and also motivate them to evolve from low epistemological maturity levels to a more mature epistemological level where they are more likely to search and develop their own knowledge. Additionally, our proposal to implement the activity of modelling the phenomenon of temperature inversion together with the development of experiments, may help students develop research interests and skills leading to a deeper understanding of science topics at next semesters. Therefore, it opens a window of possibilities for faculty to propose more challenging activities that might facilitate facing the great issues that society have related to the 2030 Agenda of Sustainable Development Goals.

\begin{acknowledgments}
O.I.G.P. acknowledges to Arath Mar\'in-Ram\'irez for the literature collecting process on the previous studies in harmonic oscillator in the education context. 
R.R.M. is grateful to his students for agreeing to share their evaluations anonymously for the academic analysis of this work.
B.M.R.L acknowledges fruitful discussion and support from Benjamin Raziel Jaramillo \'Avila.
O.I.G.P., R.R.M. and B.M.R.L acknowledge financial support from Tecnologico de Monterrey through the project grant NOVUS-2020-308.
The authors declare no competing financial interest and thank Tecnologico de Monterrey Writing Lab and TecLabs for financial support on the open access article.
\end{acknowledgments}

%

\appendix

\section{Questionnaire} \label{app:Questionnaire}

1. When preparing your answers on the thermal inversion experiment, did you consider several methods to answer the questions or did you just consider one method to find the solution? Please explain the reasoning that led to the methods that you used.

2. Did you ask for help to answer the questions? If so, who did you ask (for example, instructors, classmates, friends, family, or tutors)?

3. When answering the exercises or preparing the report, did you check any additional source of information (such as videos, books, or tutorials) besides the data given to you by the instructors? If so, which sources did you check?

4. Do you think that experimental activities based on the application of a physical concept (such as the activity on damped movement) helps improve your understanding on thermal inversion? Please explain the reasoning behind your answer.

5. Do you think that the questions you ask yourself during the experiment only have one correct answer, or that there could be more than one correct answer depending on your interpretation of certain variables and data? Please explain the reasoning behind your answer.

6. Did you feel that you had sufficient background knowledge to understand the thermal inversion experiment and to correctly answer the argumentative exam? If not, please explain which subjects would have helped you better understand the experiment.
	
\end{document}